\documentclass[12pt,preprint]{aastex}

\slugcomment{Accepted for Publication in ApJ April 8 2008}

\shorttitle{HN Peg B: A Test of Models of the Transition}
\shortauthors{Leggett et al.}

\newcommand\teff{\mbox{$T_\mathrm{eff}$}}
\newcommand\fsed{\mbox{$f_\mathrm{sed}$}}
\newcommand\logg{\mbox{$\log g$}}

\begin{document}

\title{HN Peg B: A Test of Models of \\
the L to T Dwarf Transition
}


\author{S. K. Leggett\altaffilmark{1}}
\email{sleggett@gemini.edu}

\author{D. Saumon\altaffilmark{2}}

\author{Loic Albert\altaffilmark{3}}

\author{Michael. C. Cushing\altaffilmark{4}}

\author{Michael C. Liu\altaffilmark{4,5}}

\author{K. L. Luhman\altaffilmark{6}}

\author{M. S. Marley\altaffilmark{7}}

\author{J. Davy Kirkpatrick\altaffilmark{8}}

\author{Thomas L. Roellig\altaffilmark{7}}

\and

\author{K. N. Allers\altaffilmark{4}}

\altaffiltext{1}{Gemini Observatory, Northern Operations Center, 670
  N. A'ohoku Place, Hilo, HI 96720} 

\altaffiltext{2}{Los Alamos National Laboratory, PO Box 1663, MS F663, Los Alamos, NM 87545}

\altaffiltext{3}{Canada-France-Hawaii Telescope Corporation, 
65-1238 Mamalahoa Highway, Kamuela, HI 96743}

\altaffiltext{4}{Institute for Astronomy, University of Hawai'i, 2680
  Woodlawn Drive, Honolulu, HI 96822}

\altaffiltext{5}{Alfred P. Sloan Research Fellow}

\altaffiltext{6}{Department of Astronomy and Astrophysics, The Pennsylvania State University, 
University Park, PA 16802}

\altaffiltext{7}{NASA Ames Research Center, Mail Stop 245-3, Moffett
  Field, CA 94035}

\altaffiltext{8}{IPAC, California Institute of Technology, Mail Code
  100-22, 770 South Wilson Avenue, Pasadena, CA 91125}

\altaffiltext{9}{Some of the data presented herein were obtained at the
  W.M. Keck Observatory, which is operated as a scientific partnership
  among the California Institute of Technology, the University of
  California, and the National Aeronautics and Space Administration. The
  Observatory was made possible by the generous financial support of the
  W.M. Keck Foundation. Some data were also  obtained at the Canada-France-Hawaii Telescope (CFHT)
which is operated by the National Research Council of Canada, the Institut
National des Sciences de l'Univers of the Centre National de la Recherche
Scientifique of France, and the University of Hawaii.}

\begin{abstract}

  Luhman and collaborators recently discovered an early-T dwarf companion to the G0 dwarf
star HN Peg, using $Spitzer$ Infrared Array Camera (IRAC) images. 
Companionship was established on the basis 
of the common proper motion  inferred from 1998 Two Micron All Sky Survey images
and the 2004 IRAC images.  In this paper we present new near-infrared imaging data
which confirms the common proper motion of the system.  We also present new 3 - 4 $\mu$m
spectroscopy of HN Peg B, which provides tighter constraints on both the bolometric
luminosity determination and the comparison to synthetic spectra.  
New  adaptive optics imaging data are also presented, 
which shows the T dwarf to be unresolved, providing
limits on the multiplicity of the object. We use the age, distance and luminosity of the 
solar-metallicity T dwarf to determine its effective temperature and gravity, and
compare synthetic spectra with these values, and a 
range of grain properties and vertical mixing, to the observed 0.8 - 4.0 $\mu$m 
spectra and mid-infrared photometry.
We find that models with temperature and gravity appropriate for the older end of 
the age range of the system (0.5~Gyr) can do a reasonable job of fitting the data, 
but only if the photospheric condensate cloud deck is thin, and if there is significant vertical mixing in the atmosphere.
Dwarfs such as HN Peg B, with well-determined metallicity, radius, gravity and temperature 
will allow development of dynamical atmosphere models, 
leading to the solution of the puzzle of the L to T dwarf transition.

\end{abstract}


\keywords{
stars: low-mass, brown dwarfs --- stars: individual (HN Peg, HN Peg B, 2MASS J21442847+1446077)
--- binaries: visual}

\section{Introduction}

Detailed studies of brown dwarf companions to main sequence stars
contribute significantly to our understanding of these fascinating
objects (e.g., Gl 229B, Gl 570D, HD 3651; Marley et al. 1996; 
Geballe et al. 2001; Saumon et
al. 2000, 2006; Liu et al. 2007). This is because the primary star, if
well-studied, immediately gives us the brown dwarf's distance, chemical
composition and, most importantly (since brown dwarfs cool with time),
constrains its age.   Therefore the
discovery of two T dwarfs with spectral types of T2.5 and T7.5 as
companions to the nearby stars HN Peg (G0 V, 18.4 pc) and HD 3651 (K0 V,
11.1 pc), respectively, by Luhman et al. (2007, hereafter L07; HD 3651 B was 
independently discovered by Mugrauer et al. 2006) is an
exciting and important result.  Companionship of the T dwarfs was confirmed using
proper motions determined from Two Micron All Sky Survey (2MASS;
Skrutskie et al. 2006) and \textit{Spitzer Space Telescope} (Werner et
al. 2004) Infrared Array Camera (IRAC; Fazio et al. 2004) images
separated by roughly 6 years.

In this paper we present new observational data for one of the L07 T
dwarfs, the proposed companion to HN Peg, HN Peg B (or 2MASS
J21442847$+$1446077). The G0~V primary,
HN Peg (HD 206860), is a BY Draconis variable,  where the variability has
been interpreted as due to spots on the surface, and the 24.9 day period
as the rotation period of the star (Blanco, Catalano \& Marilli 1979).
It is relatively young, with an age of $0.3 \pm 0.2$ Gyr (L07, see also \S 5 below),
and has a debris disk with a radius determined from the 70~$\mu$m flux excess 
of $\sim$7~AU (Trilling et al. 2008).
The proposed companion is $43\farcs 2$ away from the primary, which 
translates to 795~AU at the distance of the primary. Recent multiplicity 
studies suggest that the frequency of brown dwarf companions at wide
separations is low (e.g. Kraus et al. 2008), however the statistical
significance of these findings is not well constrained due to the small sample sizes.  This system
adds  one more wide-separation G star and brown dwarf pair to such studies.

Here we use the new observational data to confirm that the early-T 
dwarf is indeed a companion to HN Peg, to investigate its multiplicity, 
and to refine the determination of its luminosity. 
The luminosity allows us to constrain the temperature and gravity
of the dwarf from the known age and distance of the solar-metallicity system.
Having determined the temperature and gravity, we compare the red, 
near-infrared and 3--4~$\mu$m spectra, as well as the 
IRAC 4--8~$\mu$m photometry, to synthetic spectra and photometry and present 
the results of the model comparisons. We find that the models can reproduce the
data quite well, and hence bring us closer to resolving the puzzle of
this poorly understood phase of brown dwarf evolution: the transition
 from the dusty, red in the  near-infrared,
L dwarfs, to the clear atmosphere, blue in the near-infrared, T dwarfs
(e.g. Burgasser et al. 2002, Knapp et al. 2004).

\section{New Observations}

\subsection{Near-Infrared Imaging}

\subsubsection{WIRCam Imaging}

$J$-band imaging was conducted with WIRCam at CFHT (Puget et al. 2004) 
at three different epochs (2006 September 13, 2007 May 7
and 2007 July 12, UT) using a 9-point dithering pattern of 60$\arcsec$ amplitude.
HN Peg was positioned in the corner of the North-East detector, about 85$\arcsec$ 
off the center of the mosaic. The total integration time was 9 minutes
for the first two epochs and 17 minutes for the last. Images were preprocessed
and sky subtracted at CFHT with the I`iwi pipeline\footnote{
http://www.cfht.hawaii.edu/Instruments/Imaging/WIRCam/} and
median-stacked using the Terapix software suite (sextractor - Bertin \& Arnouts
(1996), scamp - Bertin (2006) and swarp\footnote{
http://terapix.iap.fr}). The internal
astrometry at each epoch is better than 40 milli-arcsec $rms$ and the image
quality on the resulting stacks is 0$\farcs$76, 1$\farcs$02 and 0$\farcs$81 for the three
respective epochs. Using the 2MASS
point-source catalogue to fix the astrometry of each reduced image
implies $rms$ errors of 0$\farcs$25 and this is the uncertainty that we
adopt for the coordinates of HN Peg B derived from these data.

\subsubsection{SOFI Imaging}

Through a search of the data archive for the European Southern Observatory,
we found publicly available images of HN~Peg that were obtained with the
SOFI near-infrared camera on the 3.5~m New Technology Telescope at La Silla
Observatory. These data were collected on the night of 2006 June 15
through program 077.C-0704. The instrument contained one $1024\times1024$
HgCdTe Hawaii array with a plate scale of $0\farcs288$~pixel$^{-1}$.
Ten dithered 60~second exposures of HN~Peg were obtained through an $H$-band
filter. After these images were flat-fielded, registered, and combined,
the resulting image exhibited a FWHM near $1\arcsec$ for point sources.
We determined the plate solution of the combined image using coordinates
measured by 2MASS for sources which were well-detected but unsaturated, and not
obviously multiple. The $rms$ errors in the astrometry is 0$\farcs$19.

\subsubsection{NSFCAM2 Imaging}

$K$-band images were obtained of HN Peg B on 2007 September 27
(UT) using NSFCAM2 (Shure et al. 1994) at the NASA Infrared Telescope
Facility (IRTF).  Twelve 60-second dithered frames were obtained, on a
non-photometric night, with seeing around FWHM 0$\farcs$8 at $K$. The
camera field of view is 80$''$$\times$80$''$ and the pixel scale is
0$\farcs$04/pixel.  Three well-detected stars with 2MASS near-infrared
magnitudes of 14--15 (i.e. not HN Peg or HN Peg B) were
included in five of these frames and these were used to define the
astrometry.  The uncertainty in these coordinates is estimated to be $0\farcs 1$
from the standard deviation of the  values derived from the five frames, and is 
similar to the $rms$ error in the astrometric calibration of each frame.

\subsection{3.5 $\mu$m Spectroscopy}

We acquired spectra from 2.96 to 4.07~$\mu$m of HN Peg B using
the Near-InfraRed Imager and spectrograph (NIRI, Hodapp et al. 2003) on
the Gemini North Telescope. The 3 to 4~$\mu$m wavelength region includes
both the $\nu_3$ fundamental absorption band of CH$_4$ at 3.3~$\mu$m and
a bright flux peak around 4~$\mu$m. This spectral region is therefore
useful for defining the overall shape of the spectral energy
distribution and for measuring the bolometric luminosity.

Over the course of four nights (2007 July 26 and 30, August 1 and 5, UT)
we obtained a total of 4.67 hours of data, consisting of 280 60-seconds
exposures, each made up of 30 coadds of 2-second integrations (short
integration times are necessary due to the high and variable $L$-band
background).  The $L$-grism was used with the $L$ order-sorting filter.
The nights were required to be photometric with good image quality, to
maximize the flux through NIRI's 6-pixel (0$\farcs$72) slit.  The
spectral resolving power provided by this configuration is $R \equiv
\lambda/\Delta \lambda \approx$460.
 
The F0V star HD 194822 or the A1V star HD217186 were used as
calibrators.  The science target and calibrators were observed using
$\pm$3$''$ offsets along the slit in an ABBA pattern.  Flat fields were
obtained using the Gemini calibration unit, and bad pixel masks were
derived from dark frames.  Wavelength calibration was achieved using
telluric and intrinsic spectral features in the calibration stars.  IRAF
routines were used to mask bad pixels, flat field each frame, subtract
pairs of frames, and produce one to five stacked images for each night,
each of which contained a positive and negative spectrum and represented
20--40 minutes of observation.  Figaro routines were used to
extract the spectra, wavelength calibrate and flux calibrate using the
telluric standards.  The final absolute flux calibration was achieved
using the \textit{Spitzer} IRAC 3.6~$\mu$m photometry given by L07 using
the technique given in Cushing et al. (2006).  Finally the spectrum was
rebinned so that each pixel corresponds to a single resolution element.
The signal-to-noise ratio ranges from 5 at the blue end of the spectrum
(where there is little flux) to 15--20, at the brighter red end. The
spectrum is presented later, in \S 4.

\subsection{Keck Laser Guide Star Adaptive Optics Imaging}

The G0 primary, HN Peg, has been the target of radial velocity monitoring, 
but there is no evidence of planets orbiting the star 
(Fischer \& Valenti 2005; K$\bar{\rm o}$nig et al. 2006).
Hence the system seems to be composed of the star and its 
brown dwarf companion, with a separation of 795~AU.

We searched for companions to HN Peg B on 2006 October 14
using the laser guide star adaptive optics (LGS AO) system
(van Dam et al. 2006, Wizinowich et~al. 2006)
of the 10-meter Keck II Telescope on Mauna Kea, Hawaii.  
Conditions were photometric with
variable seeing.  We used the facility infrared camera NIRC2 with its narrow
field-of-view camera, which produces an image scale of
$9.963\pm0.011$~mas/pixel (Pravdo et al. 2006)
and a $10\farcs 2 \times 10\farcs 2$ field of view.  
The LGS provided the wavefront reference source 
(equivalent to a $V\approx9.6$~mag star) for AO correction, 
with the exception of tip-tilt motion.  
Tip-tilt aberrations and quasi-static changes in the image of
the LGS were measured
contemporaneously with a lower-bandwidth wavefront sensor
monitoring the $R=12.7$~mag field star USNO-B1.0~214428+14465
(Monet et al. 2003), located 44\arcsec\ away from HN Peg B.
Technical difficulties with this second wavefront sensor led to
somewhat degraded image quality in the inner $\approx 0\farcs 3$
radius compared to typical data.  

We obtained a series of dithered images, offsetting the telescope by a
few arcseconds, with a total integration time of 420~seconds.  We used
the $K$-band filter from the Mauna Kea Observatories (MKO) filter
consortium (Simons \& Tokunaga 2002; Tokunaga et al. 2002).
The images were reduced
in a standard fashion.  We constructed flat fields from the
differences of images of the telescope dome interior with and without
continuum lamp illumination.  Master sky frame were created from
the median average of the bias-subtracted, flat-fielded images and
subtracted from the individual images.  Images were registered and
stacked to form a final mosaic, with a full-width at half-maximum of
$0\farcs 06$ and a Strehl ratio of 0.34.  No companions were clearly
detected in a $6\arcsec \times 6\arcsec$ region, corresponding to 110~AU,  
centered on HN Peg B.

We determined limits on any companions by first convolving
the final mosaic with an analytical representation of the PSF's radial
profile, modeled as the sum of multiple gaussians.  We then measured
the standard deviation in concentric annuli centered on HN Peg B, 
normalized by the peak flux of the targets, and took
10$\sigma$ as the flux ratio limits for any companions.  These limits
were verified with implantation of fake companions into the image
using translated and scaled versions of the science target.  Inside
about $0\farcs 25$ in radius, the complex structure of the PSF is the
main noise source.  From about 0.25--1.0\arcsec, the noise primarily
arises from shot noise of the PSF halo and at larger separations from
both shot noise of the sky emission and detector read noise.

Figure 1 presents the  companion detection limits.  We use
the Liu et al. (2006) polynomial fits for $K$-band absolute
magnitude as a function of spectral type to convert the brightness limits into
spectral types for field ultracool dwarfs.
Evolutionary models calculated by one of us (DS), using cloud-free atmospheres,  
were used to convert the $K$ limits into  \teff\ and mass, given the age range of
0.1--0.5~Gyr for the HN Peg system.
The observations show that there is no companion warmer than $\sim$ 500~K or
more massive than $\sim$ 10 M$_{Jupiter}$ at 0$\farcs$3 to 6$\arcsec$, 
or 5.5 to 110~AU from HN Peg B. The luminosity of the system rules out the 
presence a similar-mass companion (\S 5 and \S 6),
i.e. a companion with  mass $>$ 10--20 M$_{Jupiter}$ or \teff\ $\gtrsim$ 1000~K.  
Thus any companion to HN Peg B must be less massive than $\sim$ 10 M$_{Jupiter}$
and cooler than 500--700~K (depending on the exact age of the system).

Recent high-spatial resolution imaging of brown dwarfs has found that the binary frequency amongst T2 -
T4.5 dwarfs is unusually high, with most early-T dwarfs consisting of a similar mass, but very different color,
pair, made up of a
late-L or very early-T and a mid- to late-T dwarf (e.g. Burgasser et al. 2006b, Liu et al.  2006).
This finding is appealing as it reduces the $J$-band brightening seen for early-T dwarfs, making the
L to T transition easier to model. HN Peg B appears to be one of the rare T2.5 dwarfs which is not
composed of a similar-mass pair of dwarfs.

\section{Astrometry of the HN Peg System}

\subsection{Proper Motion Determination}

We used the WIRCam, SOFI and NSFCAM2 imaging data described above, 
as well as the $Spitzer$ IRAC images obtained for the L07 program on 2004 July 10,
to determine accurate coordinates for the proposed
companion to HN Peg at multiple epochs. The results are given in Table 1.
In all cases the dwarf was well-detected and 
2MASS stars were used to refine the coordinate systems. 

In the case of the IRAC data, coordinates could be determined from the four IRAC
channels to 0$\farcs$1$-$0$\farcs$22 in Right Ascension and
0$\farcs$1$-$0$\farcs$14 in Declination.  The uncertainty in these
coordinates is estimated from the standard deviation of the 
values derived from the four frames, and is similar to the $rms$ error in the astrometric
calibration of each frame.  

Figure 2 shows the difference between the 2004-, 2006- and 2007-epoch
measurements of the Right Ascension and Declination of the dwarf,
and the values that it would have were it a companion to HN
Peg. The 1998 2MASS coordinates provide the initial values.  We adopt a
proper motion for HN Peg of $231.2\pm 1.0$~mas/yr in Right Ascension and
$-113.6\pm 0.40$~mas/yr in Declination, based on a weighted average of
the Hipparcos (Perryman et al. 1997) and PPM-North (Roeser \& Bastian
1988) catalogs. The errors in the values plotted are the combined uncertainties 
in our astrometric measurements, the 2MASS reference
system, and the proper motion of HN Peg.

The NSFCAM2 astrometry differs in Right Ascension by
$\sim 0\farcs 3$ from the WIRCam results, and also disagrees with other available data.  
The NSFCAM2 field is known to suffer from distortion at the $0\farcs 1$ level
\footnote{http://irtfweb.ifa.hawaii.edu/\~nsfcam2/Distortion\%20Analysis.html}. 
It appears that this distortion, combined with the small number of stars used to 
define the astrometry, leads us to underestimate the error on this measurement.

The proper motion of the T dwarf in Declination is small and is
consistent with that of HN Peg along this axis. 
While the agreement in Right Ascension is not as good as that in
Declination, the values are consistent with 
companionship within the measurement errors, with the exception of the NSFCAM2 value.
Given the low probability of detecting a field T dwarf (see the following section), we interpret this
level of agreement as confirmation of companionship.

\subsection{Probability of Detecting a Background T Dwarf}

The probability of L07 discovering a background T dwarf near HN Peg is
low, although finite.  Metchev et al. (2008) determine the space density of
T0--T8 dwarfs to be 7.2$\times$10$^{-3}$ pc$^{-3}$.  The L07 study required a
good detection in all four IRAC bands, limiting the detection of early T
dwarfs to around 30~pc (and later T dwarfs to smaller distances).
The number of T dwarfs within a 30~pc volume is 810, 
using the Metchev et al. density value.  The field of view of IRAC is
5.2 arcminutes and thus each star observed by L07 samples 27.04
arcmin$^2$ or 1.82$\times$10$^{-7}$ of the celestial sphere.  The L07
sample size is 121 targets so we would expect to find $<$0.018  field T
dwarfs in this survey, as the 30~pc distance is an upper limit.
Assuming Poisson statistics, the probability of detecting one field 
T dwarf given an expectation rate of $<$0.018 is $<$2\%.
However the probability of finding such a field T dwarf with 
a similar proper motion to HN Peg (as described in the previous section) is negligibly small.

\section{The Spectral Energy Distribution and Luminosity of HN Peg B}

L07 determined the spectral type of HN Peg B to be T2.5$\pm$0.5
by comparing its near-infrared spectrum to the T dwarf spectral
templates presented by Burgasser et al. (2006a).  Using the L07 spectrum, 
we computed the values of the spectral indices
defined in Burgasser et al.  and find spectral types ranging from T2 to
T4.5, with an average of T3$\pm$1 (a significant range in type is not
uncommon for L to T transition dwarfs, see for example Table 9 in Knapp
et al. 2004).  This spectral type is in good agreement with the spectral
type derived by the direct comparison technique.
                          
Figure 3 shows the 0.8$-$4.1~$\mu$m spectrum of the Burgasser et al. T2
spectral standard SDSS J125453.90-012247.4 (hereafter SDSS 1254$-$0122)
from Cushing et al. (2005), and that for HN Peg B.  
The spectra have been scaled to demonstrate the similarity between their
spectral energy distributions (SEDs). Comparing the scaling factor to that derived
from the trigonometric parallaxes measured for SDSS 1254$-$0122  and HN Peg
(Dahn et al. 2002, Perryman et al 1997) shows that    
SDSS 1254$-$0122 is brighter than HN Peg B by a factor of 1.55. 
It has been suggested that SDSS 1254$-$0122 is multiple, 
based on its absolute magnitude (Burgasser et al. 2006b; 
Liu et al. 2006), although it remains unresolved by the \textit{Hubble Space
  Telescope} (Burgasser et al. 2006b).
The fact that HN Peg B is not brighter than SDSS 1254$-$0122 supports the
conclusion derived from the AO imaging of the dwarf,  that 
HN Peg B does not have a close companion with similar mass. 
Compared to the T2 dwarf SDSS 1254$-$0122,
HN Peg B has slightly stronger H$_2$O absorption at 1.1~$\mu$m, slightly stronger
CH$_4$ at 1.6~$\mu$m, and is brighter at 4~$\mu$m, supporting the slightly later type
of T2.5--T3.

We have determined the bolometric flux at Earth of HN Peg B 
by integrating the observed $0.8-4\,\mu$m spectrum and adding the contribution of
longer wavelengths using a synthetic spectrum. 
Initially we chose a synthetic spectrum with \teff\ $=$ 1400~K, the temperature of
SDSS 1254$-$0122 (Golimowski et al. 2004), due to the similarity in the SED.
The model flux was scaled by the observed IRAC fluxes of HN Peg B.  
After deriving the luminosity in this way, evolutionary models were used
to constrain the \teff\ and \logg\ (see the following section),
and the luminosity was rederived using the long-wavelength flux of a cooler
\teff\ $=$  1115~K model, as indicated by the evolutionary model. 
The difference in the measured luminosity is
3.5\%, less than the uncertainty in the measurement, and the change in the \teff\ 
derived from the luminosity would be $<$5~K; we have not rederived the \teff\
and \logg\ . 

The value derived for the luminosity at the Earth, using the cooler model long-wavelength
extension, is $1.66 \times 10^{-15}\,$W/m$^2$ with an estimated
uncertainty of 5\%. Adopting the HN Peg distance of 18.4 $\pm$ 0.2 pc
(Perryman et al 1997)  implies a luminosity given by 
$\log L/L_\odot=-4.76\pm 0.02$, in agreement with the value found by L07 of $-4.77\pm 0.03$.

\section{Age, Metallicity, Temperature and Gravity of HN Peg B}

L07 consider the Li, rotational, and chromospheric properties of HN Peg A to determine an age
for the HN Peg system of $0.3 \pm 0.2$ Gyr.  This agrees  with the recently derived 
gyrochronology age of 0.24 $\pm$ 0.03 Gyr (Barnes  2007).
The system  has solar metallicity: Valenti \& Fischer (2005) report [m/H]$= -0.01 \pm 0.03$ for HN Peg A.

Evolutionary models calculated by one of us (DS), show that a T dwarf aged  0.1 to 0.5 Gyr, with
our measured luminosity, has \teff\ K, \logg\ , radius $R/R_{\odot}$ and mass of 
1015~K, 4.22, 0.134, 12 M$_{Jupiter}$ at the 
younger age and 1115~K, 4.81, 0.101, 28   M$_{Jupiter}$   at the older age.  
These values are summarised in Table 2, and are consistent with the Burrows
et al (1997) and Baraffe et al (2003) evolutionary sequences shown in Figures 10 and 11 of L07,
although the temperature is slightly lower than that adopted by L07 of 1130~K.

As also noted in L07, a temperature of $\sim$1065~K is significantly cooler than that of the 
typical
field T2--T3 dwarf, which has \teff\ $\sim$ 1200--1400~K (Golimowski et al. 2004).
Cushing et al. (2008) determine \teff$=$1200--1400~K for the very spectrally similar
T2 reference dwarf SDSS J1254$-$0122 (Figure 3), from model analyses of the observed spectra,
in agreement with the value determined from luminosity
arguments by Golimowski et al. (2004) of 1425$\pm175$~K. 
However, for HN Peg B to be as hot as 1200~K, its age would need to be
$\sim$1~Gyr, in violation of the Li, rotation and chromospheric activity constraints (L07).
The young L7.5 dwarf HD 203030 B (Metchev \& Hillenbrand 2006) also has an apparently low
temperature (of 1200~K), and it has been suggested that  
the temperature of the L to T dwarf transition may be gravity dependent. 
Benchmark transition dwarfs such as HN Peg B, with well determined
age and metallicity, are clearly important for studies of the properties that
control the L to T dwarf transition.

In the following section we explore the fits of synthetic spectra and photometry
to the HN Peg B data, and investigate how well our current atmospheric models can do, given the
tight constraints on temperature, gravity and radius that the evolutionary
sequences provide. Note that if HN Peg B consists of an unresolved pair of identical
dwarfs, then the luminosity is halved and \teff\ becomes $\sim$900~K, for the age of the system.
The comparison to the spectra below shows that our models do not support such a 
low temperature, consistent with the dwarf being single, or being composed of a
significantly unequal-mass system, in agreement with the conclusions reached in \S 2.3 above.

\section{Model Comparisons}

The  model atmospheres used here self-consistently
include the formation of condensate clouds (Marley et al. 2002, 2007, and in
preparation).  The parameter \fsed\ is a measure of the efficiency of
condensate sedimentation relative to turbulent mixing (Ackerman \& Marley
2001).  Larger values of \fsed\ imply larger particles sizes, greater 
sedimentation efficiency, and thus thinner condensate clouds.
Generally, \fsed\ $=$ 2 reproduces the colors and spectra of L dwarfs well, and
\fsed\ $=$ 4 or cloud-free models reproduce the colors and spectra of
later T dwarfs well (e.g. Knapp et al. 2004, Cushing et al. 2008).

The models also include vertical transport in the atmosphere, which affects
the chemical abundance of species involving C, N and O. The extremely
stable CO and N$_2$ molecules can be dredged from deep layers into the photosphere,
enhancing the abundances of these species, and decreasing the abundance
of CH$_4$, H$_2$O and NH$_3$ (e.g. Fegley \& Lodders 1996,
Saumon et al. 2003 and 2007, Hubeny \& Burrows 2007).
This mechanism is parameterized in our models by a diffusion coefficient $K_{zz}$
cm$^2$s$^{-1}$. The larger $K_{zz}$, the greater the enhancement of CO and
N$_2$ over CH$_4$ and NH$_3$. For T dwarfs, the effect is significant 
in the mid-infrared, where the 3~$\mu$m CH$_4$ absorption band is weakened,
the 4.5~$\mu$m CO absorption band is strengthened, and the 
11~$\mu$m NH$_3$ absorption band is weakened (e.g. Leggett et al. 2007).
Values of  $\log K_{zz} = 2$ to 6, 
corresponding to mixing time scales of $\sim$ 1~h to $\sim 10$ yr in the 
atmosphere, appear to be required to reproduce the observations of T dwarfs.

\subsection{0.8--4.1~$\mu$m Spectra}

Figure 4 shows the observed 0.8--4.1~$\mu$m spectrum of HN Peg B compared to
various model spectra.
We treat the sedimentation and diffusion parameters
\fsed\ and $K_{zz}$ as independent parameters, although when full hydrodynamic
models can be calculated the grain sedimentation and replenishment will most
likely be found to be intimately connected with the vertical transport of gas
through the atmosphere. Here we use models  with \fsed\ $=$ 1, 2, 3, 3.5 and 4, and with
$K_{zz} = 0$(equilibrium), $10^2$, $10^4$ and $10^6$  (cm$^2$s$^{-1}$), and values of \teff\ / \logg\
for each end of the allowed age range, 1015/4.22 and 1115/4.81 (\S 5).

In all cases we find the spectra generated by the hotter temperature and higher 
gravity models do a better job of reproducing the data than the lower temperature
and gravity models.  Thus the age of the system appears to be at the older end of the
allowed range, close to 0.5~Gyr.  The best fitting model has \teff\ $=$ 1115~K, 
\logg\ $=$ 4.81, \fsed\ $=$ 3.5 and $K_{zz} = 10^4$ cm$^2$s$^{-1}$ (red line in Figure 4). 
The cloud decks in this dwarf must therefore be relatively thin.  
Figure 4 also shows that significant vertical
mixing in the atmosphere is required to reproduce 
the observed depth of the 3~$\mu$m CH$_4$ feature.

\subsection{4--8~$\mu$m Photometry}

Figure 5 compares the observed IRAC absolute magnitudes of HN Peg B to those calculated
by our models.  The models match the data quite well, except for the  4.49~$\mu$m band,
where the discrepancy is $\sim$0.4 magnitudes. This is puzzling, given how well the model 
reproduces the 3--4~$\mu$m spectrum (Figure 4), and the 5.73~$\mu$m magnitude. At these
wavelengths the flux emerges from above the condensate clouds, and so our modelling of the cloud decks
is not the source of the error. The problem either lies in the temperature-pressure profile 
---  the region around 0.5 to 1.0 bar appears to be too hot --- or in an opacity source that is missing or too weak in the models.

Nevertheless, the modelled photometry reproduces the data quite well, and supports the 
conclusions drawn from the comparison to the shorter-wavelength spectrum ---
high sedimentation, thinner cloud deck are required, as well as significant vertical mixing.
The best fitting model for the
0.8--4$\mu$m spectrum is also the best choice to fit the IRAC photometry: \teff\ $=$ 1115~K, \logg\ $=$ 4.81, \fsed\ $=$ 3.5 and $K_{zz} = 10^4$ cm$^2$s$^{-1}$.

\section{Conclusions}

The discovery by L07 of an early-T dwarf companion to a solar-like star offers an important testbed
for models of the complex and poorly understood transition from the dusty L dwarfs to the clear T dwarfs. 
The new imaging data presented here improves the determination of the proper motion of this T dwarf,
and confirms that the brown dwarf has common proper motion with
the star, and hence is indeed a companion.  

Our new high spatial-resolution imaging data,
together with the model comparisons to the 0.8--4.1~$\mu$m spectra, shows
that the dwarf is most likely single, although it is possible that a much fainter, undetected,
$< 10$ M$_{Jupiter}$ brown dwarf lies within 100 AU of the early-T dwarf.

Comparison of the observed and  synthetic spectra generated by our models indicates that 
the HN Peg system is at the older end of the allowed age range, with an age $\sim$0.5~Gyr.  
Together with the measured
luminosity, this implies that HN Peg B has \teff\ $=$ 1115~K, \logg\ $=$ 4.8, radius 0.10
R$_{\odot}$, and mass 28 M$_{Jupiter}$. 

Comparison of the observed and  synthetic 0.8--8.0~$\mu$m photometry and spectra also indicates that the
condensate cloud decks in the photosphere of this T dwarf must be relatively thin, as we determine
\fsed\ $= 3.5$. Vertical mixing appears to be significant, as implied by the relative
weakness of the 3~$\mu$m CH$_4$ absorption band, which we fit with $K_{zz} = 10^4$ cm$^2$s$^{-1}$.
In the future, objects like this, with well-determined age, radius, mass, gravity and temperature,
should allow development and testing of  dynamical atmosphere models
that use a more physical approach to treat clouds and mixing.  In the longer term, such models should
provide the solution to the puzzle of the L to T dwarf transition.

\acknowledgments

This work uses observations obtained at the
Gemini Observatory through Program GN-2007B-Q-22. Gemini Observatory is
operated by the Association of Universities for Research in Astronomy,
Inc., under a cooperative agreement with the NSF on behalf of the Gemini
partnership: the National Science Foundation (United States), the
Science and Technology Facilities Council (United Kingdom), the National
Research Council (Canada), CONICYT (Chile), the Australian Research
Council (Australia), CNPq (Brazil) and CONICET (Argentina).
We are very grateful to John Rayner for the NSFCAM2 observations.  
We would also like to thank the observers and queue coordinators who carried service
observations at CFHT (programs 06BD94, 07AD98 and 07AD84).
We gratefully acknowledge the Keck LGS AO team for their exceptional
efforts in bringing the LGS AO system to fruition.  It is a pleasure
to thank Randy Campbell, Jim Lyke, Cindy Wilburn, Joel Aycock, and the
Keck Observatory staff for assistance with the observations.
This work is based and supported (in part) on
observations made with the \textit{Spitzer Space Telescope}, which is
operated by the Jet Propulsion Laboratory, California Institute of
Technology under a contract with NASA. 
This publication makes use of data from the Two Micron All Sky Survey, which
is a joint project of the University of Massachusetts and the Infrared
Processing and Analysis Center, and funded by the National Aeronautics
and Space Administration and the National Science Foundation, the SIMBAD
database, operated at CDS, Strasbourg, France, and NASA's Astrophysics
Data System Bibliographic Services.  
SKL's research is supported by Gemini Observatory. 
MCL and KLA acknowledge support for this work from NSF grants AST-0407441 and
AST-0507833 and an Alfred P. Sloan Research Fellowship.
%


{\it Facilities:} \facility{Gemini:Gillett (NIRI)}, \facility{IRTF (SpeX)}, \facility{CFHT ()},
\facility{Keck II Telescope (LGS AO, NIRC2)}

\clearpage

\begin{deluxetable}{rrrrrl}
\tablecaption{Astrometry for HN Peg B  (2MASS~J21442847$+$1446077)}
\tablehead{
\colhead{Epoch} & \colhead{Right Ascension} &  \colhead{$\sigma$(RA)} & 
\colhead{Declination} & \colhead{$\sigma$(Dec.)} & \colhead{Survey or}\\
\colhead{Year} & \colhead{HH:MM:SS.SSS} & \colhead{arcsec} & \colhead{DD:MM:SS.SS}
& \colhead{arcsec} & \colhead{Instrument}\\
} 
\startdata
1998.734 & 21:44:28.472 & 0.07 & $+$14:46:07.80 & 0.07 & 2MASS \\
2004.441 & 21:44:28.548 & 0.22 & $+$14:46:07.06 & 0.14 & IRAC \\
2006.540 & 21:44:28.595 & 0.19 & $+$14:46:06.86 & 0.19 & SOFI \\
2006.699 & 21:44:28.582 & 0.25 & $+$14:46:06.81 & 0.25 & WIRCam \\
2007.348 & 21:44:28.601 & 0.25 & $+$14:46:06.73 & 0.25 & WIRCam \\
2007.529 & 21:44:28.599 & 0.25 & $+$14:46:06.73 & 0.25 & WIRCam \\
2007.742 & 21:44:28.581 & 0.10 & $+$14:46:06.79 & 0.10 & NSFCAM2 \\ 
\enddata
\end{deluxetable}

\begin{deluxetable}{rrrrr}
\tablewidth{200pt}
\tablecaption{Luminosity-Constrained Parameters for HN Peg B}
\tablehead{
\colhead{Age} & \colhead{\teff\ } & \colhead{\logg\ } &  
\colhead{Radius } & \colhead{Mass } \\
\colhead{Gyr} & \colhead{K} & \colhead{} &  
\colhead{$R_{\odot}$} & \colhead{$M_J$} \\
} 
\startdata
0.1 & 1015 & 4.22 & 0.134 & 12\\
0.5 & 1115 & 4.81 & 0.101 & 28\\
\enddata
\end{deluxetable}

\clearpage

\begin{figure}
\includegraphics[angle=-90,scale=.60]{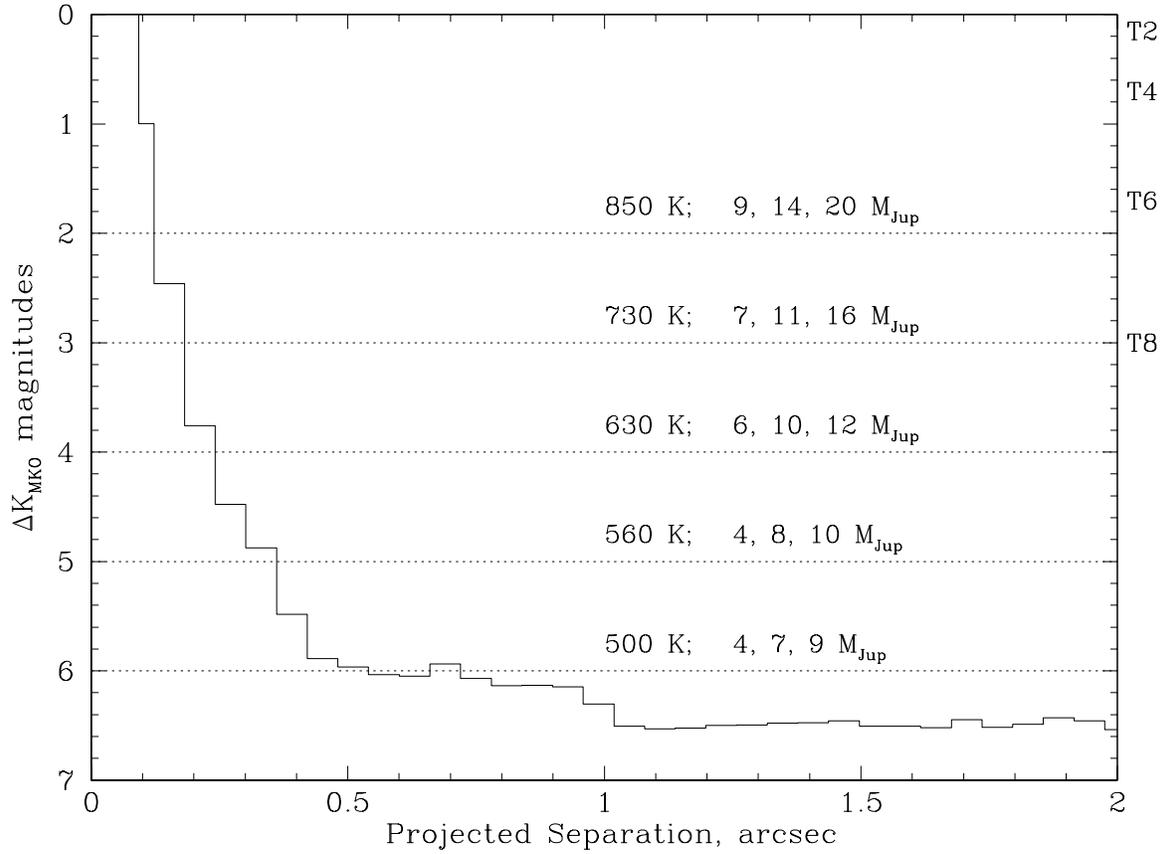}
  \caption{Limits on multiplicity of HN Peg B, derived from our adaptive optics imaging. Spectral types corresponding to the $K$ magnitude 
of a companion to HN Peg B are shown along the right axis, and  effective temperatures and masses of any companion are shown
along the dotted lines, for ages 0.1, 0.3 and 0.5 Gyr. 
 \label{fig1}}
\end{figure}

\clearpage

\begin{figure}
\includegraphics[angle=-90,scale=.60]{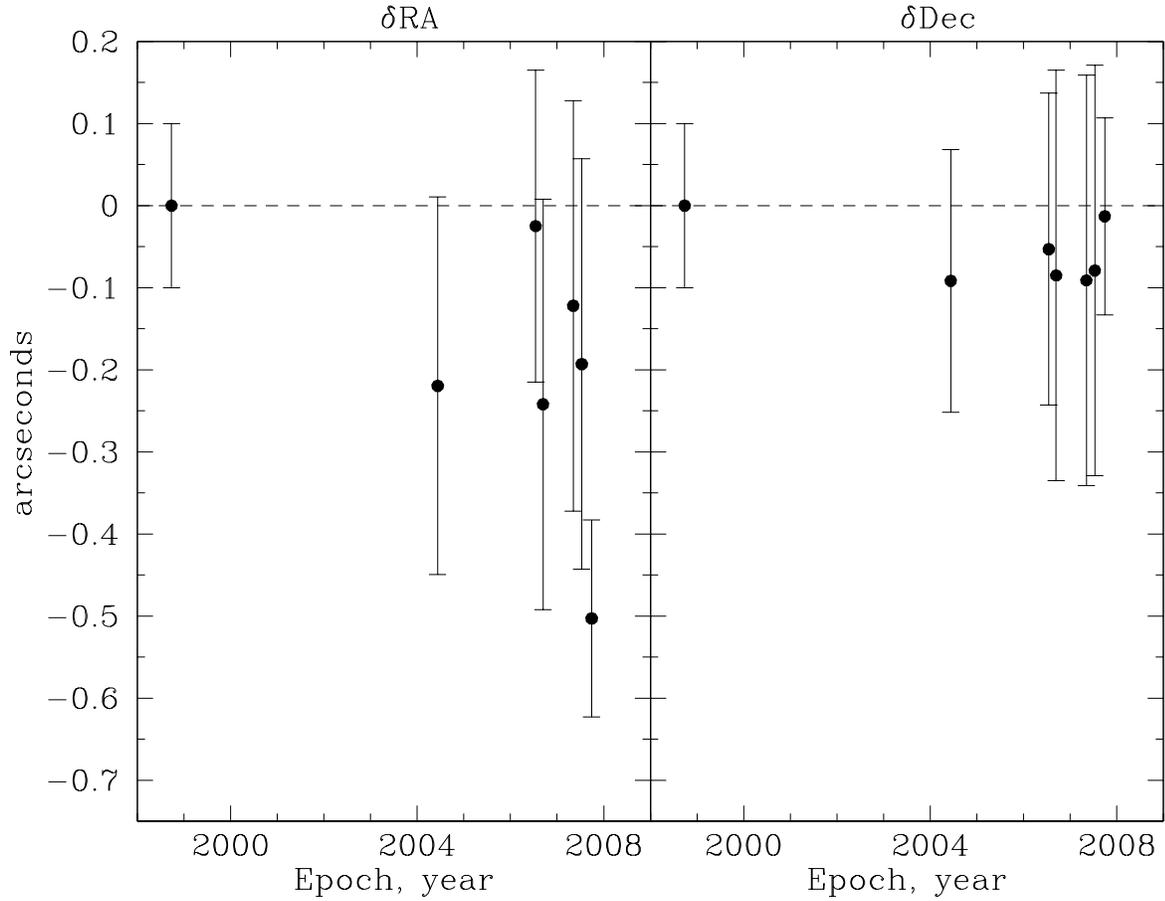}
  \caption{The difference between the positions of the proposed companion to HN Peg at various 
epochs, and its 1998.7 2MASS position, updated assuming common proper motion 
with HN Peg. The astrometry derives from images obtained with IRAC (2004.4), SOFI 
(2006.5), WIRCam (2006.7, 2007.3, 2007.5) and NSFCAM2 (2007.7).
 \label{fig2}}
\end{figure}

\clearpage

\begin{figure}
\includegraphics[angle=-90,scale=.60]{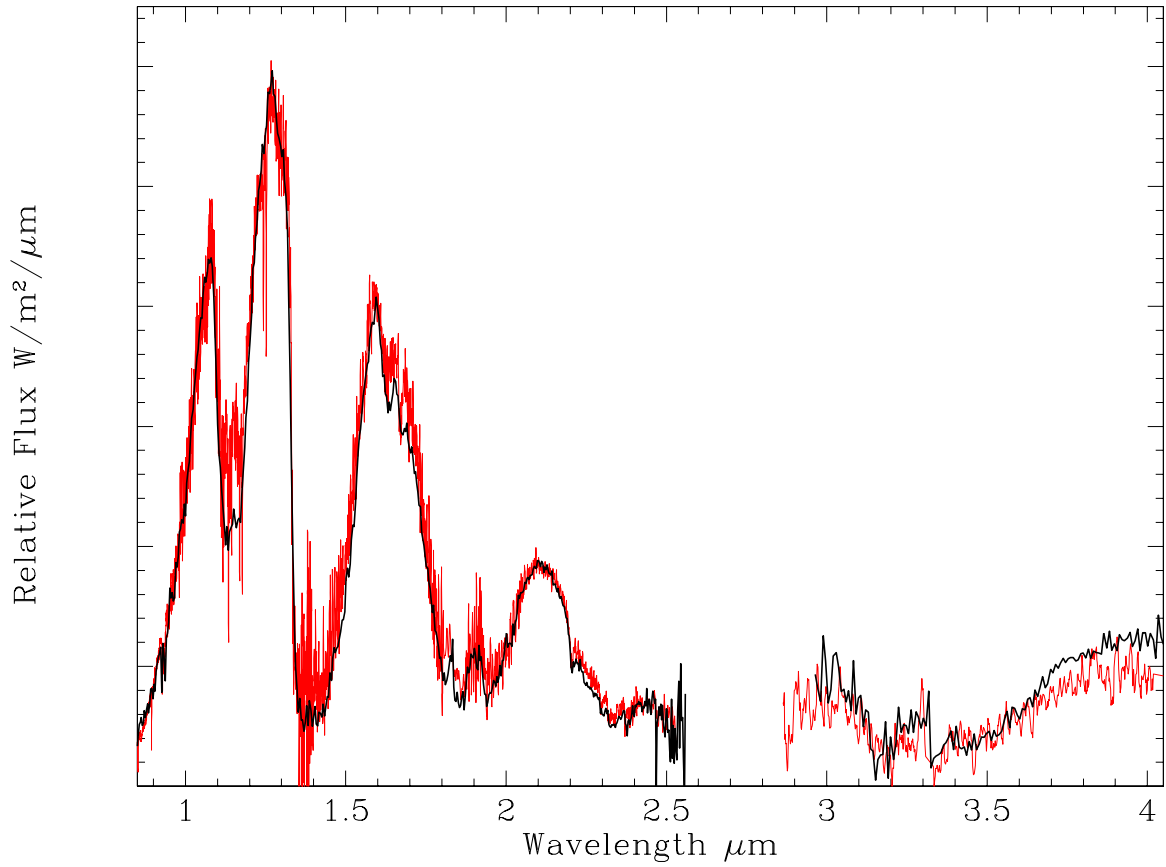}
  \caption{Observed spectra for the T2 template SDSS J1254$-$0122
    (red, Cushing et al. 2005) and HN Peg B (black, L07 and this
    work). 
  \label{fig3}}
\end{figure}

\clearpage

\begin{figure}
\includegraphics[angle=0,scale=.60]{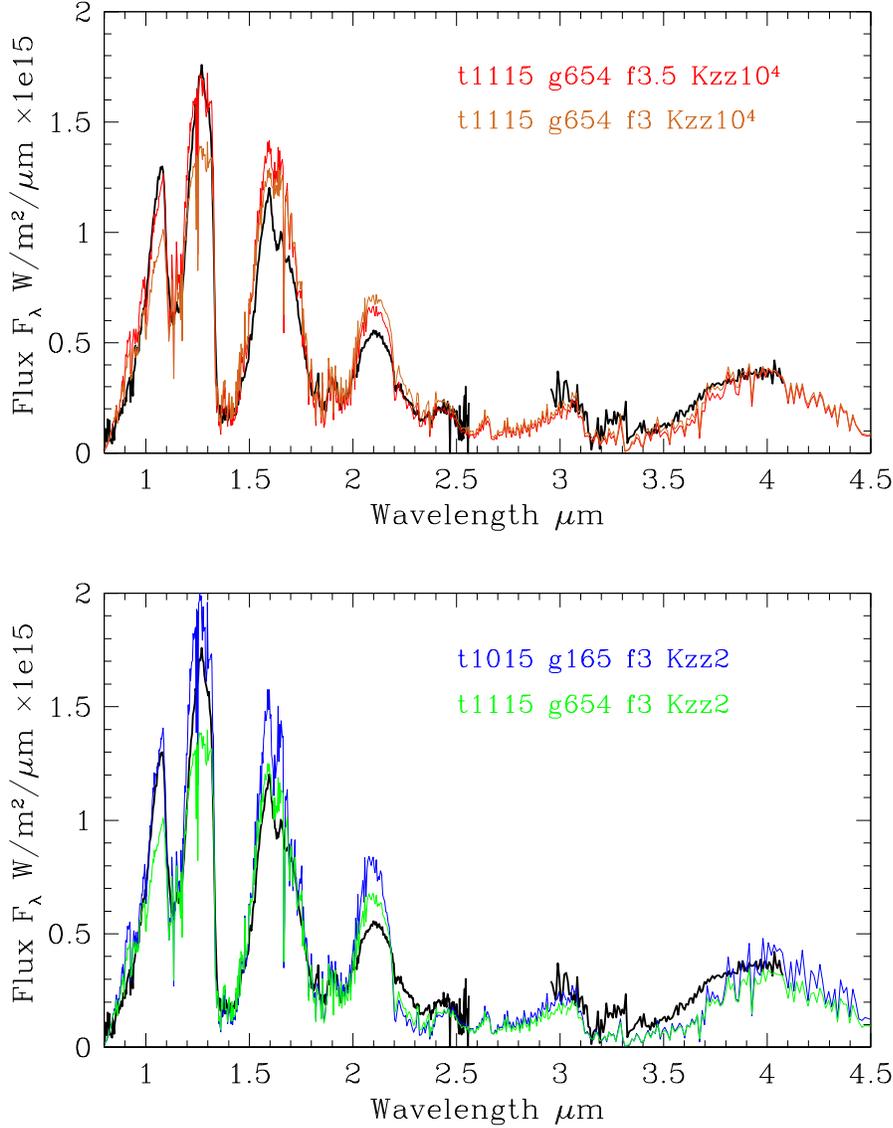}
\caption{The observed spectrum of HN Peg B (heavy black line) is
compared to synthetic spectra with various \teff\ , gravity, sedimentation
efficiency \fsed\ and vertical mixing diffusion coefficient $K_{zz}$
parameters, as indicated in the legends.  
The model fluxes have been scaled to flux at the Earth using the known distance to
HN Peg and the HN Peg B radii calculated by evolutionary models.
The top panel shows our 
best-fitting model (red line) together with a similar model with
thicker cloud decks (brown line). The bottom panel shows the
best-fitting cooler model (blue line), together with a warmer model
spectrum demonstrating the effect of less vertical transport 
(green line, compare to brown line in top panel).
\label{fig4}}
\end{figure}

\clearpage

\begin{figure}
\includegraphics[angle=-90,scale=.60]{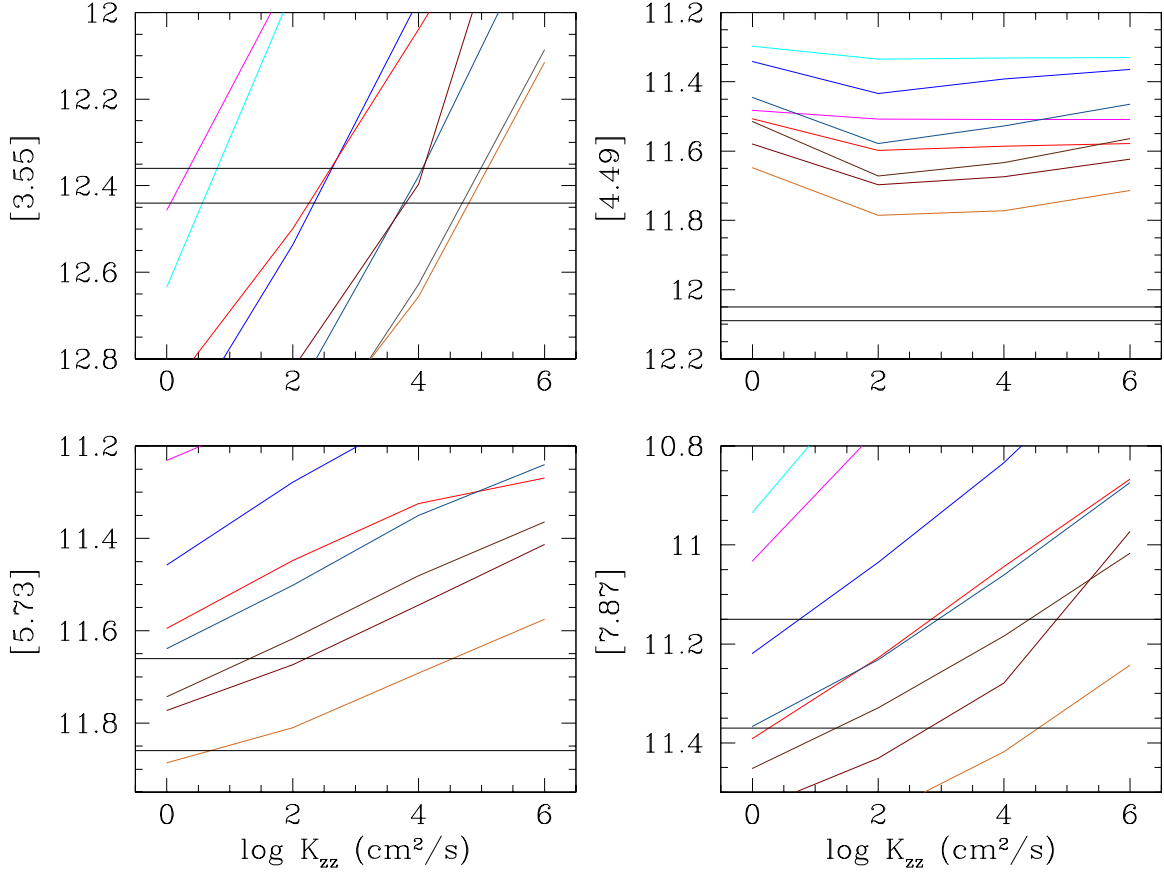}
\caption{The absolute IRAC magnitudes for HN Peg B (range indicated by
horizontal black lines), compared to synthetic photometry from our models
as a function of vertical mixing diffusion coefficient $K_{zz}$.
For ease of plotting $K_{zz}=0$ has been plotted as log~$K_{zz}=0$.
The model values cluster in pairs of 1015~K (cyan to dark blue to grey) and 1115~K
(pink to dark red to brown), with sedimentation parameter \fsed\ $=$ 1, 2, 3 and 4
from left to right for the [3.55] panel, and top to bottom in the other panels. 
\label{fig5}}
\end{figure}

\end{document}